\documentclass[aps,pra,reprint,a4paper,groupedaddress,showpacs]{revtex4-1}
\usepackage{graphics}
\usepackage{amsmath}
\newcommand{\sech}{\mathrm{sech}}

\begin{document}

\title{Nonlinear propagation of an optical speckle field}

\author{Stanislav Derevyanko}
\email{s.derevyanko@aston.ac.uk}
\affiliation{Nonlinearity and Complexity Research Group, Aston University,  Aston Triangle, Birmingham B4 7ET, UK}

\author{Eran Small}
\affiliation{Department of Physics of Complex Systems, Weizmann Institute of Science, Rehovot 76100, Israel}

\date{\today}

\begin{abstract}
We provide a theoretical explanation of the results on the intensity distributions and correlation functions obtained from a random beam speckle field in nonlinear bulk waveguides reported in the recent publication Y. Bromberg et.al., Nat. Photonics \textbf{4}, 721 (2010). We study both focusing and defocusing case and in the limit of small speckle size (short correlated disordered beam) provide analytical asymptotes for the intensity probability distributions at the output facet. Additionally we provide a simple relation between the speckle sizes at the input and output of a focusing nonlinear waveguide. The results are of practical significance for the nonlinear Hanbury Brown and Twiss interferometry both in optical waveguides and Bose-Einstein condensates.
\end{abstract}


\maketitle

\section{Introduction}
\label{sec:intro}
Extreme events play an important part in many areas of physics. In turbulence they determine the non-Gaissian statistics of the tails of the probability distribution functions for the properties of random flow \cite{inst} and in the linear theory of random wave localization the momenta non-selfaveraging quantities, like e.g. wave transmissivity are determined by rare non-typical realizations rather than the typical (localized) ones \cite{lgp,sgm99}.

In the context of the nonlinear optics the study of the extreme events has recently drawn attention in the context of the \textit{optical rogue waves} \cite{rogue} - emerging dynamical objects of very high amplitude and short lifetime. Closely related topics are extreme statistics in Raman amplification \cite{hpf11} and the \textit{supercontinuum generation} \cite{dge08}. The appearance of the rogue waves in optical fibers is not necessarily a nonlinear effect and can be observed in the linear regime as well \cite{vt11}. Finally the statistics of rare events (extreme outages) also determines the probability of errors in fiber optical communications with distributed amplifier spontaneous emission \cite{mec}.

Most of the applications above pertain to the field of nonlinear fiber optics. Here we study the emergence of the high power optical pulses in the context of the nonlinear Hanbury Brown and Twiss (HBT) interferometry. The goal of this paper is to explain theoretically recent experimental results reported in Ref.\cite{blss10} on the non-exponential distribution of the intensity distributions of disordered optical field propagating in nonlinear bulk waveguides. The linear HBT method was first proposed in 1950s in astronomy as a means of measuring a size of a distance light source (e.g. a star) by measuring the \textit{intensity correlation radius} of the received light \cite{hbt}. The latter is the the typical size of an optical \textit{speckle} (i.e. bright area) observed when multiple waves emitted from a thermal source interfere constructively \cite{Goodman}. The problem allows classical treatment and when the propagation is linear one can infer that at a distance $Z$ from a source of diameter $L$ the size of the speckle $S$ is given by a simple relation $S= Z \lambda/L$, where $\lambda$ is the wavelength of emitted light \cite{blss10}.

Ref.\cite{blss10} was the first publication to our knowledge where the HBT interferometry in \textit{nonlinear light} propagation and the resulting speckle distribution were studied in bulk AlGaAs waveguides. The system is effectively described by the Nonlinear Schrodinger Equation (NLSE) which paves the way to application of the nonlinear HBT interferometry not only in the field of optics but weakly interacting Bose-Einstein condensates as well \cite{BEC}. Two principal findings of Ref. \cite{blss10} were that the tail of the intensity distribution, $P(I)$, of the speckled field is non-exponential (unlike in a linear diffraction \cite{Goodman}) and the intensity correlation radius (speckle size) depends on the magnitude of the focusing nonlinearity. The first observation signifies the fact that the statistics of the optical filed after the nonlinear propagation is no longer Gaussian (or, equivalently, the amplitude of the field no longer follows the Rayleigh distribution). This effect is also known to occur in linear systems when a linear wave propagates through a disordered media \cite{sk91}. Here however we deal with a different type of setup where not only is the non-linearity present but also disorder is only in the initial conditions (incident beam) and not in the media itself. For the focusing nonlinearity in an important 2D configuration of the system (commonly known as 1+1 geometry) this phenomenon was correctly attributed by the authors of Ref.\cite{blss10} to generation of \textit{bright spatial soliton} beamlets \cite{KivAgr} that now play the role of observed speckles. The tails of the intensity distribution are determined by extremely rare events when an extremely high power (and narrow width) soliton is born from a random beam of a finite waist and intensity. Our paper seeks to explain theoretically the profile and shape of the tails of the intensity probability density functions (PDFs) in the 1+1 geometry using the inverse scattering technique (IST) for the NLSE \cite{mnpz}. We first construct a semi-empirical theory of a HBT interferometer in the high-power regime where the field is dominated by its soliton component. This theory is later corroborated by the results obtained analytically from the IST in the limit of short correlated source field. As for the dependence of the intensity correlation radius on the magnitude of nonlinearity observed in \cite{blss10} this is of course to be expected in the nonlinear system. Here we also supply a simple theoretical result for the average speckle size $S$ in the same short-correlated source limit, and derive theoretically the linear scaling of its inverse, $S^{-1}$, with the average intensity $I_0$ of the source (or, equivalently, with the nonlinear coefficient). We also provide theoretical results in the case of de-focusing nonlinearity where no bright solitons are observed. In all cases we perform full numerical simulations using the parameters close to those used in the experimental setup of Ref.\cite{blss10} to confirm our analytical predictions.

Note that the propagation of incoherent fields in nonlinear optics has been studied previously in various contexts (see e.g. Refs. \cite{moti}). However most of the systems considered usually rely on more complicated non-local or non-instantaneous models and the quantities calculated are usually not particularly relevant for the HBT problem. The local 1+1 NLSE is the simplest model where one can get both a clear physical understanding of how the nonlinearity affects HBT interferometry and obtain some analytical results. Similar problem statement occurs in the context of fiber optics Ref.\cite{td08}, where the formation of bright solitons from a disordered input was employed to illustrate the concept of \textit{soliton discriminator} as a means of nonlinear optical regeneration. However in \cite{td08} as opposed to the current paper an opposite regime was addressed where the emergence of a soliton from a disordered output has a relatively small probability and neither the intensity distribution nor the correlation function of the output were studied.

\section{The model}

The dynamics of beam propagation along the direction of $z$ axis in the presence of diffraction and nonlinearity is given by the nonlinear Schrodinger equation (NLSE) \cite{KivAgr}:
\begin{equation}
\frac{\partial E}{\partial z} = \frac{i}{2\beta_0}\,\frac{\partial^2 E}{\partial x^2} +i\, \frac{n_2}{n_0} \beta_0 \, |E|^2 E
\label{NLSE0}
\end{equation}
where $E$ is the electrical field, $x$ is the spatial transverse coordinate, $\beta_0=(2\pi/\lambda) \, n_0$ is the propagation constant, $n_0$ is the linear refractive index and $n_2$ is the nonlinear coefficient of the medium. Here we will consider both attractive ($n_2>0$) and repulsive ($n_2 < 0$) cases.
We will also be using dimensionless soliton units $\xi =x/L$, where $L$ is some characteristic width, $\tau = z/ L_D$ where $L_d=L^2\, \beta_0$ is the diffraction length (Rayleigh range) and $u$ is the dimensionless field $E/\sqrt{\tilde I}$, $\tilde I =(|n_2|\,\beta_0\,L_d/n_0)^{-1}$. Then in new dimensionless units we will have
\begin{equation}
\frac{\partial u}{\partial \tau} = \frac{i}{2}\,\frac{\partial^2 u}{\partial \xi^2} +\frac{n_2}{|n_2|}\,i\, |u|^2 u
\label{NLSE}
\end{equation}

When choosing the model for random disordered input we opt for the form which mimics as close as possible the experimental setup and simulation data from Ref.\cite{blss10}. The physical input is defined by two parameters: the initial speckle size $S_0$ and the aperture $L$ (we also pick the latter as the normalization width for the soliton units above). In our numerical simulations the spatial resolution $\delta x$ is determined according to $S_0$ to insure sufficient sampling and the width of computational domain $L'$ is set large enough to prevent folding during the propagation. The input disordered field is modeled in Fourier space by $N$ random low frequency modes, where $N=L'/S_0$. As for the complex amplitudes of these random modes, $a_n$, we assume that these form a set of of independent identical random variables each having a uniformly distributed phase and the amplitude sampled from a distribution with the average intensity $a^2$. This pattern is then filtered by the finite aperture $L$ to manifest the disordered field at the input facet of the nonlinear waveguide. As the number of modes is large the central limit theorem applies so that the field above, $E_0(x)$, can be considered Gaussian with zero mean and the correlation function:
\begin{equation}
\label{correl-init}
\langle E_0(x) \,E_0^*(x')\rangle =a^2 \,\frac{\sin \left[\pi(x-x')N/L'\right]}{\sin \left[\pi(x-x')/L'\right]}
\end{equation}
The field is of course non-zero only in the aperture window $[-L/2,L/2]$ so the formula above applies only to this region. The averaged initial intensity is then $I_0 = \langle |E_0(x)|^2 \rangle = a^2\,N$.
Also when $S_0$ is the smallest length scale in the problem we will be using the delta-correlated approximation when the r.h.s. of Eq.(\ref{correl-init}) is substituted with $2\tilde{D} \delta(x-x')$ where $\tilde{D}=L' a^2/2=S_0 I_0/2$. This makes further analytical treatment possible in some cases and we will often use it in the paper.

As for the intensity distribution, $P(I)$, since the initial field is Gaussian the statistics of the propagated field in the linear case ($n_2=0$) will remain Gaussian. As the intensity is the modulus squared of the complex Gaussian variate its normalized value $I/\langle I \rangle$ has an exponential distribution, the fact which is well known in the linear theory of speckle spectra \cite{Goodman}. Our main task in the subsequent sections will be to determine the modified shapes of the intensity distribution $P(I)$ in the presence of nonlinearity. Of particular interest is the high intensity tail of this distribution determined by rare fluctuations leading to the field bursts.

\section{The phenomenological approach}
\label{sec:focus}
Let us consider the case of focusing ($n_2>0$) NLSE. Then it is known that given enough initial power an arbitrary initial condition $E_0(x)$ evolves into the combination of hyperbolic secant constituents (each corresponding to an individual bright soliton) and quasilinear radiation \cite{mnpz}. Here we will adopt a phenomenological description of the intensity distribution and the correlation properties of the output field based on the prescribed form of the solution as the sum of statistically independent soliton pulse shapes with prescribed statistical properties.

A single soliton solution of (\ref{NLSE}) in soliton units is given by:
\begin{eqnarray}
u_s(\tau,\xi)&=& 2\eta \,\sech \left[4\eta \, v \,\tau+2\eta\,(\xi-\xi_0)\right] \nonumber \\
 & \times & \exp \left[-i(2\,v\,\xi+2\tau(v^2-\eta^2)-\phi)\right]
\label{one-sol}
\end{eqnarray}
where parameters $\eta$ and $v$ are related to the soliton's amplitude, $A$, and `velocity' (i.e. the angle of incidence, $\theta$), while the parameters $\xi_0$ and $\phi$ are the soliton's initial position and global phase. The total power of an individual soliton (or rather its transversal part in the $x$-plane) $P=\int |E|^2 d x$ is simply proportional to its amplitude: $P=4\,\eta \, \tilde{I}\, L = 4\eta n_0/|n_2| \beta_0^2 L$.

The justification for this approach is presented in the following chapter \ref{sec:IST} where more rigorous IST based analysis is performed. It is these soliton constituents that contribute to the tails of intensity distribution $P(I)$ as the linear radiation quickly disperses away from the aperture. Let us now assume the regime where the density of solitons is not too high so that the average minimal distance between the solitons is larger than the average width of a soliton -- a regime that can be called an asymptotically free regime. Then each soliton contributes independently into the intensity distribution (the interference effects are neglected) and we may consider a contribution from each individual soliton separately.
For a single soliton solution of the NLSE (\ref{NLSE0}) in the real world units the intensity at a given point $x$ is given by:
\begin{equation}
I(x;\eta,v,x_0,\tau)=4\, \eta^2\, \tilde{I} \,\sech^2 \,\left[2\,\eta\,\left(\frac{x-x_0}{L}+2 v \tau \right)\right],
\label{intens}
\end{equation}
where $2\eta$ is the amplitude of the soliton in the dimensionless soliton units, $2v$ is its velocity and $x_0$ is the intial position of the soliton center (in $\mu$m). As the input is random the amplitude, velocity and the initial position of the soliton are also random variables. As for their joint probability density function $P(v,x_0,\eta)$ we shall make an assumption which is supported both by numerical simulations and the following Zakharov-Shabat eigenvalue analysis (see below). Namely we assume that soliton velocity parameter $v$ is independent from the other variables and its distribution is uniform over a symmetric interval $[-\Delta v/2, \Delta v/2]$. This immediately means that for a soliton emitted from the origin the position shift due to the fluctuating velocity is also a uniformly distributed random variable in the interval $[-\tilde \Delta/2,\tilde \Delta /2]$, where $\tilde \Delta = 2 \Delta v \tau L$ and $\tau=Z/L_d$ is the propagation distance in soliton units.

The conditional probability density of having the value of intensity in the vicinity of $I$ given the amplitude of the soliton, $2\,\eta$, is then given by
\begin{equation}
P(I|\eta;x,\tau)=\left\langle \frac{1}{\Delta v} \, \int_{-\Delta v/2}^{\Delta v/2} \, \delta\left[I- I(x;\eta,v,x_0,\tau) \right] \, dv \right \rangle
\label{P-I-eta0}
\end{equation}
where $I(x;\eta,v,x_0,\tau)$ is given by Eq.(\ref{intens}) and the angular brackets denote additional averaging with respect to the marginal PDF of the initial positions $P(x_0|\eta)$.

In order to perform the averaging analytically we will only consider the high intensity tails of the PDF when the typical soliton width, $L/2\eta$ is much more narrow than the width of position distribution, $\tilde \Delta $.
Additionally we will assume that the propagation distance $\tau$ is large enough so $\tilde \Delta>>L$, and therefore the fluctuations in the soliton initial position, $x_0$ , are negligible when compared to these due to the random velocity $2v$. After all these assumptions the result of integration (\ref{P-I-eta0}) can be presented as
\begin{equation}
P(I|\eta;x)=\frac{L}{\tilde \Delta} \,\frac{\sqrt{\tilde{I}}}{I\,\sqrt{4\,\eta^2\,\tilde{I}-I}}\,\theta\left[4\,\eta^2\,\tilde{I}-I\right], \, |x| < \tilde \Delta/2
\label{P-I-eta}
\end{equation}
where $\eta \gg L/2 \tilde \Delta$ and $I \gg 16 \eta^2 \tilde{I} \,\exp[-8\eta \Delta v \tau]$.

If the number of produced solitons is $n>1$ then it is relatively straightforward to derive the average minimal distance between the neighbouring solitons which is given by $\tilde \Delta/(n^2+1)$. As we have assumed that the width of each soliton is much more narrow than the average minimal inter-soliton distance this implies that the condition for the amplitude is in fact $\eta \gg L(n^2+1)/2 \tilde \Delta$.

Finally to get the marginal intensity distribution $P(I)$ we need to average Eq.(\ref{P-I-eta}) over all realizations of soliton amplitude $2\eta$. Assuming that the marginal PDF $P_\eta(\eta)$ is known and is the same for all realizations with different soliton numbers the result reads:
\begin{equation}
P(I)= \langle n \rangle \, \frac{1}{4 \Delta v \, \tau}\,\frac{1}{I}\,\intop_{0}^{\infty}\,P_\eta(\sqrt{I/4\tilde{I}}\,\cosh z) \,dz,
\label{pdf-onesol}
\end{equation}
where $I \gg (\tilde{I}/(2\Delta v \,\tau)^2)\,(\langle n \rangle^2+1)^2$ and factor $\langle n \rangle$ takes into account that for each realization all $n$ solitons contribute equally and independently into the intensity and we neglect the effects of interference and soliton collisions (i.e. overlap).

\section{IST based analysis}
\label{sec:IST}
Many properties of the evolving solution of NLSE (\ref{NLSE}), including the number of emerging solitons, their amplitudes, energies and velocities can be established by means of so called Zakharov-Shabat spectral problem (ZSSP) \cite{mnpz}:
\begin{equation} \label{ZSSP}
\begin{split}
i\,\frac{\partial \psi_1}{\partial \xi} + u \psi_2 = \zeta \psi_1 \,
,
\\
-u^* \psi_1 -i\,\frac{\partial \psi_2}{\partial \xi}  =  \zeta
\psi_2 \, .
\end{split}
\end{equation}
Here the complex initial field $u(0,\xi)$ plays the role of the ``potential'' while the complex eigenvalues $\zeta$ can have both discrete and continuous values. It is the discrete spectrum $\zeta_n=v_n+i\,\eta_n$ which determines the soliton part of the solution and in the case of a single soliton solution the parameters $v$ and $\eta$ are are exactly the ones featured in Eq. (\ref{one-sol}).

\begin{table}[h]
\caption{\label{tab:1} The main parameters of the simulations for the focusing case}
\begin{tabular}{|l|l|}
  \hline \hline
  Parameter& Value \\ \hline
  The size of the aperture, $L$, $\mu$m  &  50  \\ \hline
  The thickness of the slab, $d$, $\mu$ m & 1.5 \\ \hline
  The size of the computational domain, $L'$, $\mu$m &  4096 \\ \hline
  The total number of points, $M'$ & $8192$ \\ \hline
  The number of points resolving the aperture, $M$ & 100 \\ \hline
  The total number of random modes, $N$ & 2048  \\ \hline
  The maximum propagation distance, $z$, $\mu$m & 8000  \\ \hline
  Linear refraction index, $n_0$ & 3.3  \\ \hline
  Propagation constant, $\beta_0$, $\mu$m$^{-1}$ &  12.61 \\ \hline
  The nonlinear coefficient, $n_2$, cm$^2$/GW & $1.67 \times 10^{-4}$ \\ \hline
  Initial correlation length, $S_0$, $\mu$m & 2 \\ \hline
  The diffraction length, $L_d$, mm & $33.9$\\ \hline
  Normalization intensity, $\tilde{I}$, GW/cm$^2$ & $0.05$  \\ \hline
  Window used for collecting histograms, $\Delta$, $\mu$m & 1024 \\ \hline
  \hline
\end{tabular}
\end{table}

\begin{figure}[h!]
\scalebox{1.0}{\includegraphics{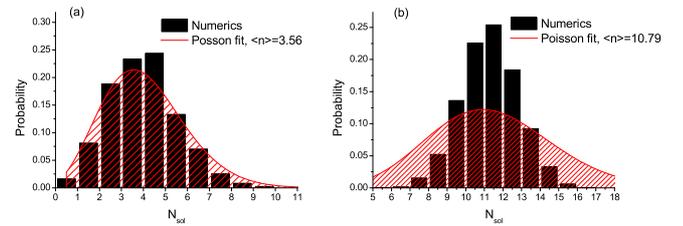}}
  \caption{\label{fig:numbers} (Color online) The distribution of the number of emerging solitons for low intensity (a) and high intensity (b) regimes}
\end{figure}

We will start with the numerical Monte Carlo simulations for the number of emerging solitons as well as their amplitudes and phases. In the Monte-Carlo simulations we took 4000 runs with the parameters of random Zakharov-Shabat potential given in Table \ref{tab:1}. Those were chosen to be close to experimental values of Ref.\cite{blss10}. We performed two runs with different values of initial average power, $P=a^2 \,N\, d\,L$. The first one corresponds to peak power of $P=$1kW and the second to $P=5$kW and we will refer to these as ``low power'' and ``high power'' runs correspondingly (keeping in mind that these labels correspond to the power of the initial disordered filed). Other runs with different values of the input parameters (like the initial power and the correlation length) were also performed but their results were qualitatively the same as in the ``high power'' or ``low power'' runs so to keep the presentation simple and illustrative we only report the results for these two. Because the input is random the number of emerging solitons will fluctuate around the mean. In Fig.\ref{fig:numbers} we present a distribution of the number of emerging solitons for the values of parameters given in Table \ref{tab:1} together with the corresponding Poisson fits.

One can clearly see that the number of soliton does approximately follow the Poisson distribution for low intensity (small number of solitons) but this approximation breaks down for high power run when the number of solitons is higher. Similar results were reported earlier in Ref. \cite{td08} in the context of the nonlinear fiber optics.

\begin{figure}[h!]
\scalebox{1.0}{\includegraphics{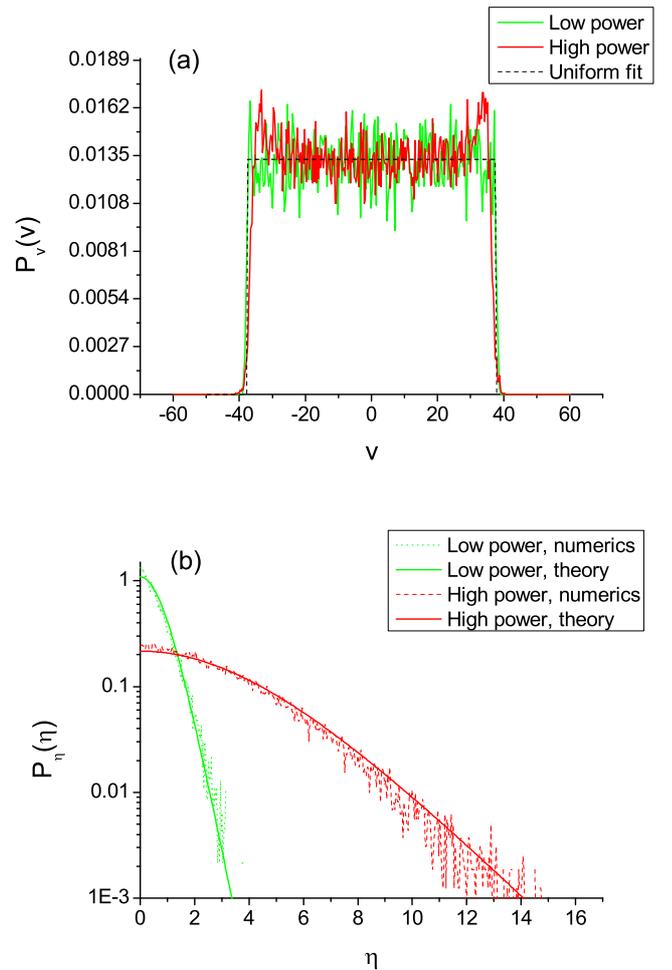}}
  \caption{(Color online)\label{fig:etaV} The marginal PDFs of the real (a) and imaginary (b) parts of the ZSSP eigenvalues.}
\end{figure}

Next, shown in Fig.\ref{fig:etaV} (as dashed and dotted lines) are the numerical marginal PDFs for the real and imaginary parts of ZS eigenvalues $P_v(v)$ and $P_\eta(\eta)$ which are just scaled distributions for soliton final position and amplitude. A separate Monte-Carlo run (not shown) has demonstrated that the numerical joint distribution $P(v,\eta)$ is indeed separable so the marginal distributions are sufficient. One can see that the PDF for the real part $P_v(v)$ is close to uniform with the width $\Delta v=75.8$ which supports the assumptions made earlier in section \ref{sec:focus}. Indeed, assuming that solitons are created in the area localized by the relative small aperture size $L$ it follows that the uniform distribution of the real parts of the eigenvalues with support $\Delta v$ yields the uniform distribution of soliton position at the output facet with the support $\tilde \Delta = 2 \Delta _v (L/L_d) z = 1788\mu$m for the values of parameters given in Table \ref{tab:1}. Moreover our numerical data also shows that the support of the distribution $\Delta$ does not depend on the average power $P$, but rather solely on the input correlation length $S_0$, where, as expected, shorter input correlation distances yield broader distributions for $v$.

In the limit of delta-correlated initial filed the results above can be confirmed analytically. Indeed one can formally define a 2D density of states (eigenvalues) of the non-Hermitian ZSSP (\ref{ZSSP}) as
\begin{equation}
\rho(v,\zeta)= \frac{1}{L}\, \sum_n \langle \delta(v-v_n) \,\delta(\eta-\eta_n) \rangle
\label{density}
\end{equation}
where summation is performed over all discrete eigenvalues for each realization and the averaging is performed over all possible realization. It is clear that the density of states $\rho(v,\zeta)$ is (up to a normalization factor) the probability density of having a level in the vicinity of point $(\xi,\eta)$. Therefore if one knows the density of states it is possible in principle to determine the desired level distribution. In Ref. \cite{km08} this quantity was obtained analytically in the limit of the Stratonovich delta-correlated potential when the support of the potential $L$, is large. The result reads:
\begin{equation}
\rho(v,\eta)=\frac{1}{\pi D} \,\frac{(\eta/D)\,\coth(\eta/D)-1}{\sinh^2(\eta/D)}, \quad D\equiv \frac{S_0}{L}\,\frac{I_0}{2\tilde I}.
\label{Greeks}
\end{equation}
One can show \cite{d11} that in the strict mathematical limit of the Stratonovich white noise the result above is inapplicable but it holds for any physical process with a symmetric field correlation function of finite but small radius,  like e.g. (\ref{correl-init}) when $S_0$ is much less than the aperture length. One immediately notices that the quantity $\rho(v,\eta)$ does not depend on the real part of the eigenvalue, $v$. Therefore the total number of states with given imaginary part, $\eta$ is infinite, i.e. the probability density function $P(v,\eta)$ is not normalizable in the $v$-direction. This is again the consequence of idealized nature of the white noise and for systems with finite correlation radius all the quantities in question are of course finite. This is indeed confirmed by the numerical results discussed above - as we can see the marginal PDF $P_v(v)$ is almost flat, it does not depend on the value of the initial power and its support diverges in the white noise limit. Thus the analytical result, Eq.(\ref{Greeks}) in the short-correlated limit explains both the separability of the eigenvalue distribution and the flat marginal distribution $P(v)$ observed numerically. One can now immediately derive the analytic expression for the normalized marginal probability $P_\eta(\eta)$:
\begin{equation}
P_\eta(\eta)=\frac{2}{D} \,\frac{(\eta/D)\,\coth(\eta/D)-1}{\sinh^2(\eta/D)}
\label{P-eta}
\end{equation}
For the specific parameters of our numerical simulations last formula in Eq.(\ref{Greeks}) yields $D=0.61$ for the low power run and $D=3.08$ for the high power run. The corresponding analytical curves (\ref{P-eta}) are plotted as solid lines in Fig.(\ref{fig:etaV})b and one can observe a rather good agreement with the numerics.

\begin{figure}[h!]
\scalebox{1.0}{\includegraphics{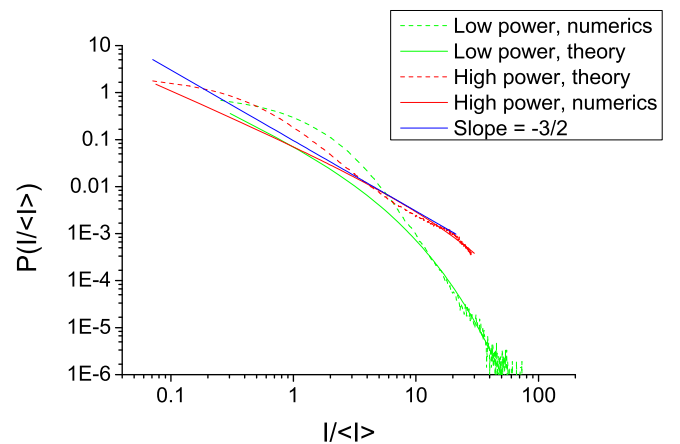}}
  \caption{(Color online) The PDF of the field intensity obtained from the direct numerical simulations of Eq.(\ref{NLSE}) and the one reconstructed from the IST theory via formula (\ref{P-est}). The average intensity was $\langle I \rangle = 3.28\times 10^{-2}$GW/cm$^2$ (low power) and $\langle I \rangle = 13.3\times 10^{-2}$GW/cm$^2$ (high power).}
  \label{fig:Pest}
\end{figure}

Plugging the amplitude PDF  (\ref{P-eta}) into the expression (\ref{pdf-onesol}) one obtains
\begin{equation}
\begin{split}
P_{est}(I)&= \langle n \rangle \,\frac{L}{2\tilde \Delta D}\,\frac{1}{I}\, f\left(\sqrt{\frac{I}{4\tilde{I} D^2}}\right),\\
f(x) & =  2 \intop_0^\infty \frac{x\,\cosh z\,\coth(x\cosh z)-1}{\sinh^2(x\cosh z)}\, dx
\end{split}
\label{P-est}
\end{equation}
For the large values of argument the $x \gg 1$ we have $f(x) \approx 8 x K_1(2x)$ and we obtain the asymptote $P(I) \propto I^{-3/4}\,\exp[-(I/D^2\tilde{I})^{1/2}]$ as the high intensity tail of the distribution. It turns out that Eq.(\ref{P-est}) provides a remarkably good approximation of the tails of the intensity PDF - see Fig.\ref{fig:Pest} where we compare it with the results of the direct Monte Carlo simulations of the NLSE propagation (again 4000 realizations were used).
One can see that our analytical result works rather well for $I >10 \langle I \rangle = 0.328$GW/cm$^2$ (low power limit) and $1.33$GW/cm$^2$ (high power limit). The discrepancies at low intensities are due to the fact that i) at low intensity the contribution of the radiation (completely ignored in our semi-analytical scheme) becomes non-negligible and ii) the dilute soliton gas approximation is not valid for very low power (and hence very wide) solitons that overlap and interfere significantly which violates the assumptions used in deriving Eq.(\ref{P-est}). For the overlapping solitons the phase interference becomes an important effect diminishing the soliton contribution into the intensity which explains why the analytical result overestimates the probability of low-intensity events. One can also observe that the numerical PDF for both high and lower power values has an interesting structure with the inflection point. This inflection point corresponds to the crossover between the regime of well separated high-intensity pulses and that of broad interfering low-intensity solitons. Finally we plot a $-3/2$ slope line as a reference. In Ref. \cite{blss10} it was suggested that this slope is rather universal which would imply some universal power-law tails. Our results show that it is not so. While it does work well as the best fit in the region around the inflection point in the high-power case it is well off the mark in the low-power case. Also for the high power case one can clearly see a crossover to the exponential tail as predicted by Eq.(\ref{P-est}).

\begin{figure}[t!]
\scalebox{1.0}{\includegraphics{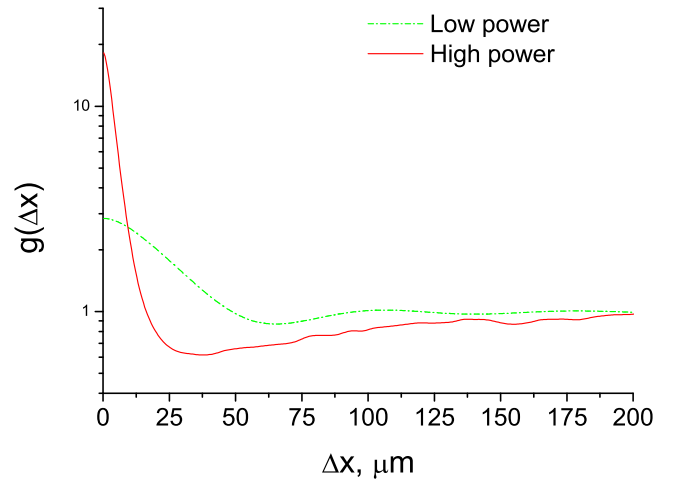}}
  \caption{\label{fig:corr} (Color online) The intensity correlation function obtained from the direct numerical simulations of Eq.(\ref{NLSE}).}
\end{figure}

Following Ref.\cite{blss10} we can also introduce a normalized intensity autocorrelation function as
\begin{equation}
\label{eq-g}
g(\Delta x) = \frac{\int \,\langle I(x) \,I(x+\Delta x) \rangle \, dx}{\int \,\langle I(x) \rangle \,\langle I(x+\Delta x) \rangle \, dx}
\end{equation}
where $I(x)=|E(x)|^2$ is the fluctuating intensity of the beam. For a linear medium it can be shown that $g(0)=2$ and then it falls to $g(\infty)=1$ over a characteristic length scale  - the intensity correlation length $S$ (also called the ``speckle size").
In Fig. \ref{fig:corr} we plot the correlation function $g(\Delta x)$ obtained from the same numerical run as the the other statistics. If we compare the limiting values of the intensity correlation function with the results of Appendix \ref{sec:corr} which were obtained using only soliton component of the solution one can see that the limiting value $g(\infty)$ is very close to unity while theory gives the value $\langle n^2\rangle /\langle n \rangle^2 -1/\langle n \rangle \approx 0.92$ (for both high and low power runs) which is close. As for the opposite limit, one can see from Fig.(\ref{fig:corr}) that the limiting values $g(0)= \langle I^2 \rangle /\langle I \rangle^2 \approx 2.85$ (low power) and $\approx 18.5$ (high power) are far less than the predictions of Eq.(\ref{fluct-n}) (where the moments of $\eta$ were taken from distribution (\ref{P-eta})), $g(0) \approx 17$ (low power) and $g(0) \approx 26.5$ (high power).
This discrepancy can be easily explained when one looks at Fig. (\ref{fig:Pest}) where it is clear that the first two moments of the intensity determining $g(0)$ are formed by the region $I \sim \langle I \rangle$ where the nonsoliton part of the radiation is important and also soliton pulses are wide enough to cover most of the sampling region. However the analytical prediction of the Appendix A for the correlation length itself, $S=L/2\sigma_\eta=L/2D \approx 40 \mu$m (low power) and $\approx 8 \mu$m (high power) holds remarkably well. If we recall the definition of the parameter $D$ we get (in the short-correlated limit) a simple relation between the observed correlation length, $S$, the initial correlation length, $S_0$, and average initial intensity, $I_0$:
\begin{equation}
S=\frac{L^2 \,\tilde I}{S_0 \, I_0}
\label{scaling}
\end{equation}
So the observed correlation length is inversely proportional to the initial correlation length and the initial average intensity (or power). The latter fact confirms the experimental results of Ref. \cite{blss10}. Also the correlation length in this regime does not depend on the propagation distance, $\tau$, (i.e. is saturated) which together with the scaling law $g(0) \propto \tau$ (see Eq.(\ref{fluct-n})) coincide with the results obtained in \cite{blss10} by qualitative considerations.

\section{Defocusing case}

Let us now turn attention to the de-focusing case where $n_2<0$. The experimental results of \cite{blss10} show qualitatively different behavior of both the correlation function and the intensity PDF. In particular the correlation strength $g(0)$ goes down with the intensity of the initial speckle field (unlike in the focusing case) and the intensity distribution also look markedly different. Here we will again employ the inverse scattering method to study the resulting intensity PDF. Similar problem for the defocusing case was studied previously in \cite{bkk87} using asymptotic far field expansion developed by Manakov \cite{mnpz,m74}. The ZSSP for focusing and defocusing case differ only by a sign of the potential in the second equation so that in the defocusing case one will have
\begin{equation} \label{ZSSP1}
\begin{split}
i\,\frac{\partial \varphi_1}{\partial \xi} + u \varphi_2 = \zeta \varphi_1 \,
,
\\
u^* \varphi_1 -i\,\frac{\partial \varphi_2}{\partial \xi}  =  \zeta
\varphi_2 \, .
\end{split}
\end{equation}
If we assume zero boundary conditions at infinity for the defocusing NLSE no bright (or dark) solitons are formed and the far field is formed solely by dispersive waves.

It is known that asymptotically at large $z$ the field intensity $I(x,z)$ in the real world units is given by the following formula \cite{mnpz,m74,bkk87}:
\begin{equation}
I(x,z)= \frac{\tilde{I} \, L_D}{2\pi z} \,\ln \left| a\left(-\frac{L_D}{2L}\,\frac{x}{z}\right)\right|^2 \label{Manakov}
\end{equation}
where $a(\zeta)$ is determined via the particular solution of (\ref{ZSSP1}) subject to the following boundary conditions \cite{mnpz}:
\begin{equation}
\phi(0;\zeta)= \left( \begin{array}{c} 1 \\ 0 \end{array}\right) \quad \phi(1;\zeta)= \left( \begin{array}{c} a(\zeta)\,e^{-i \zeta} \\  b(\zeta) e^{i\,\zeta} \end{array}\right)
\label{BC}
\end{equation}
Here the spectral parameter $\zeta$ is real and we have assumed for definiteness that the initial support of the pulse is $[0,L]$ in the real-world units. Coefficients $a(\zeta)$ and $b(\zeta)$ are called the first and second Jost coefficients respectively and satisfy the condition $|a|^2-|b|^2=1$. Using the invariant imbedding approach already developed for the focusing case (see e.g. \cite{dp08}) we can introduce the functions $a(\zeta;\xi)=\varphi_1(\xi)e^{i\zeta\,\xi}$, $b(\zeta;\xi)=\varphi_2(\xi)e^{-i\zeta\,\xi}$, for which we will have the following system of equations:
\begin{equation}
\label{Jost}
\begin{split}
\frac{\partial a(\zeta,\xi)}{\partial \xi} & = i\,b(\zeta,\xi)\,e^{2i\,\zeta \xi} \,u(\xi), \quad a(\zeta;0)=1 \\
\frac{\partial b(\zeta,\xi)}{\partial \xi} & = -i\,a(\zeta,\xi)\,e^{-2i\,\zeta \xi} \,u^*(\xi),  \quad b(\zeta;0)=0
\end{split}
\end{equation}
The Jost coefficients are recovered as $a(\zeta)=a(\zeta,1)$ and $b(\zeta)=b(\zeta,1)$.
Note that as $u(\xi)$ may be considered Gaussian with the the correlation function given by (\ref{correl-init}) (in the real world units) the phase factor $\exp(2 i\zeta \xi)$ can be absorbed into the definition of the random field $u(\xi)$ so that the statistics of both Jost coefficients becomes independent on the spectral parameter $\zeta$. From (\ref{Manakov}) it immediately follows that asymptotically the values of the field intensity become uncorrelated, i.e. the correlation function $g(\Delta x)$ tends to unity across the traverse dimension of the beam as long as the distance $z$ is large. Let us parameterize $|a|$ and $|b|$ as $\cosh \chi$ and $\sinh \chi$ respectively and introduce the phase difference between the two Jost coefficients: $\varphi = \mathrm{Arg}[a] - \mathrm{Arg}[b]$. Then for these two real quantities one obtains a system of equations:
\begin{equation}
\label{Jost1}
\begin{split}
\frac{d \chi}{d \xi} & = - \mathrm{Im} [ e^{-i\,\varphi} \, u(\xi)], \quad \chi(0)=0  \\
\frac{d \varphi}{d \xi} & = 2 \,\coth 2\chi   \, \mathrm{Re} [ e^{-i\,\varphi} \, u(\xi)],
\end{split}
\end{equation}
where the value of the initial phase $\varphi(0)$ is chosen so that the derivative $\varphi'(0)$ is finite (see e.g. \cite{dp08}). In the delta-correlated limit we obtain from the system (\ref{Jost1})(treated in the Stratonovich sense) the following Fokker-Planck equation for the joint PDF $P(\chi,\varphi;\xi)$ \cite{r}
\begin{equation}
\frac{\partial P}{\partial \xi} = -D \frac{\partial }{\partial \chi} \left[ \coth (2\chi) \,P \right] +\frac{D}{2}\,\frac{\partial^2 P}{\partial \chi^2} +2D\,\coth^2(2\chi)\,\frac{\partial^2 P}{\partial \varphi^2}
\label{FPE-marginal}
\end{equation}
According to (\ref{Manakov}),(\ref{BC}) the statistics of the intensity is determined by the statistics of the quantity $\ln \cosh \chi$ at $\xi=1$ so we can integrate out the dependence on the angular variable $\varphi$ using the periodic boundary conditions and make a substitution $P(\chi;\xi)=Y(\chi;\xi)\sinh 2\chi$ for the resulting marginal distribution of the random variable $\chi$. The resulting equation reads:
\begin{equation}
\begin{split}
\frac{\partial Y}{\partial \xi} & = \frac{(D/2)}{\sinh 2\chi} \, \frac{\partial}{\partial \chi} \, \left[ \sinh (2\chi) \,\frac{\partial Y}{\partial \chi}\right], \\
 & Y(\chi;0)  =\delta(\chi)/\sinh (2\chi)
\end{split}
\label{FPE-chi}
\end{equation}
This equation is known in the theory of stochastic processes \cite{lgp,d08} and it has the solution in quadratures:
\begin{equation}
Y(\chi;1)=\sqrt{\frac{1}{\pi D^3}}\,e^{-D/2} \,\intop_{\chi}^\infty \, \frac{\chi'\,\exp(-\chi'^2/2D)}{\sqrt{\cosh(2\chi')-\cosh(2\chi)}} \, d \chi'
\label{sol-FPE}
\end{equation}
Finally for the intensity PDF one has the following expression:
\begin{widetext}
\begin{equation}
P(I)= \frac{2\pi z}{\tilde{I} L_D} \,\exp \left[\frac{2\pi z}{L_D}\,\frac{I}{\tilde{I}}\right] \,\ Y \left[ \frac{\pi z}{L_D}\,\frac{I}{\tilde{I}} + \ln \left(1+\sqrt{1-\exp\left(-2\,\frac{\pi z}{L_D}\,\frac{I}{\tilde{I}}\right)}\right);1 \right]
\label{P-I}
\end{equation}
\end{widetext}
An interesting feature of the distribution above is that it is self-similar in the propagation distance, $z$, i.e. the PDF of the quantity $y \equiv zI/\tilde I L_D$ is universal and depends only on the disorder level $D$. The tail of this PDF is Gaussian
\begin{equation}
P(y)= \frac{\sqrt{y}}{D} \, e^{-D/2}\,e^{-y-y^2/2D} \quad y \gg 1
\label{P-I1}
\end{equation}

Also as mentioned before in this far field limit the field values at the different points are uncorrelated so that $g(\Delta x) \equiv 1$, for $\Delta x \geq S_0$.

To test the analytical result above we again choose the model parameters close to those considered in experiments \cite{blss10}. Namely we choose ethanol with an absorbing dye as a de-focusing nonlinear medium at the wavelength of 552nm and assume the values of parameters given in Table \ref{tab:2}.

\begin{figure}[h!]
\includegraphics{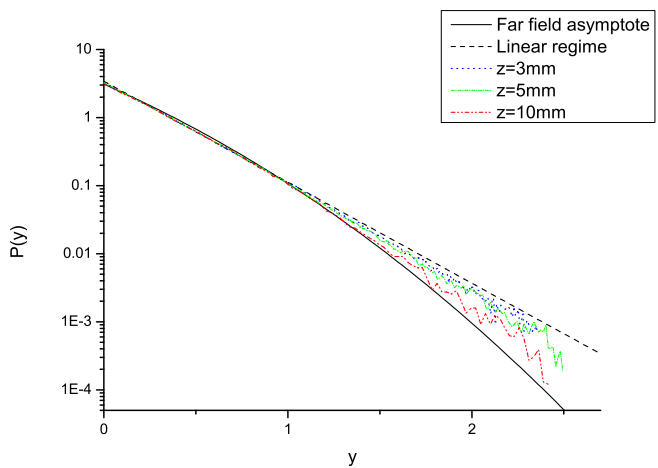}
  \caption{\label{fig:defocus} (Color online) The PDF of the self-similar variable $y$ for different values of the propagation distance.}
\end{figure}

\begin{table}[h]
\caption{\label{tab:2} The main parameters of the simulations for the defocusing case.}
\begin{tabular}{|l|l|}
  \hline \hline
  Parameter& Value \\ \hline
  The size of the aperture, $L$, $\mu$m  &  50  \\ \hline
  The thickness of the slab, $d$, $cm$ & 2 \\ \hline
  The size of the computational domain, $L'$, mm &  8.192\\ \hline
  The total number of points, $M'$ & $16384$ \\ \hline
  The number of points resolving the aperture, $M$ & 100 \\ \hline
  The total number of random modes, $N$ & 4096  \\ \hline
  Linear refraction index, $n_0$ & 1.3  \\ \hline
  Propagation constant, $\beta_0$, $\mu$m$^{-1}$ &  15.62 \\ \hline
  The nonlinear coefficient, $n_2$, cm$^2$/W & $-2.6 \times 10^{-8}$ \\ \hline
  Initial correlation length, $S_0$, $\mu$m & 2 \\ \hline
  The diffraction length, $L_d$, mm & $39.04$\\ \hline
  Normalization intensity, $\tilde{I}$, W/cm$^2$ & $82.00$  \\ \hline
  Window used for collecting histograms, $\Delta$, $\mu$m & 512 \\ \hline
  \hline
\end{tabular}
\end{table}
Although in the real experiments the pulse attenuation was quite high at the length scale considered we can still use these parameters for illustrative purposes. In our simulations we assumed a fixed averaged value of the input power, $P=6$W, which correspond to the effective nonlinear length $L_{NL}=(|n_2| \beta_0 I_0/n_0)^{-1}=5.33$mm. The simulations were performed for three values of the propagation distance $z$ comparable to the nonlinear length. The results are given in Fig.\ref{fig:defocus}

One can clearly see that depending on the propagation distance $z$ there are two regimes with two types of statistics. When the propagation distance is less than the nonlinear length, $L_{NL}$, the nonlinear term in Eq.(\ref{NLSE}) can be neglected, i.e. the propagation is linear and as mentioned before the intensity PDF is exponential. It easy to check that in terms of the self-similar variable $y$ the linear PDF is
\[
P_{lin}(y)=\frac{1}{2D} \,e^{-y/2D}
\]
When the distance exceeds the nonlinear length the far-field asymptotes (\ref{P-I}) and (\ref{P-I1}) hold and the shape of the pdf $P(y)$ becomes universal (with the Gaussian tails). The intermediate values of the propagation distance fill the gap between the two limiting distributions (as clearly seen in Fig.\ref{fig:defocus}).

\section{Conclusion}
\label{sec:concl}
To conclude, we have studied both analytically and numerically the statistics of the intensity distribution of a disordered short-correlated pulse propagating in nonlinear media under conditions close to those experimentally observed in Ref.\cite{blss10}. In the limit of the delta-correlated pulse, when the initial correlation length (speckle size) is much less than the aperture size we provide analytical expressions for the intensity distribution for both focusing and defocusing media. It turns out that the power-law tails reported in \cite{blss10} are not universal and represent an approximate fit to a transitional area followed by an exponential asymptote in the focusing case and Gaussian asymptote in the de-focusing case. For the latter a universal analytical formula for the intensity PDF is given in the regime when the propagation length (the length of the beam) is larger than the nonlinear length. Also in the focusing case a simple expression is given for the intensity correlation width, $S$ (formula (\ref{scaling})) which relates it to the initial correlation width (speckle size), $S_0$ and the average intensity of the source $I_0$ confirming the results of Ref.\cite{blss10}. This formula supplements the relation $g(0) S =2 \lambda z/L$ obtained in Ref. \cite{blss10} using quantitative arguments and thus allows one to estimate not only the linear size of the object, $L$, but also its average intensity $I_0$ and a correlation radius $S_0$ (or rather the product of the two) when the system is in a soliton regime (high power, high number of speckle beamlets).

We would like to thank Yaron Silberberg for drawing our attention to the problem and Ehud Altman for stimulating discussions. SD would like to thank the Department of Physics of Complex Systems at  Weizmann Institute of Science for its warm hospitality.

\appendix
\section{The derivation of the soliton intensity correlation function}
\label{sec:corr}
For a single soliton with the intensity profile (\ref{intens}) and the uniform position distribution it is possible to calculate the intensity correlation function $g_1$ provided that the typical soliton amplitude, $\sigma_\eta$ is much greater than the ratio $1/(4\Delta v \tau)$, i.e. all typical soliton pulse realisations (as well as the correlation function itself) have the width much less than the size of the region where soliton are eventually distributed, $\tilde \Delta$. Then one obtains
\[
\langle I_1(x) \rangle = \langle I_1(x+\Delta x) \rangle  = \frac{2\tilde{I}}{\Delta v \tau} \langle \eta \rangle \equiv I_1
\]
and
\[
\langle I_1(x) I_1(x+\Delta x) \rangle  =  \frac{16\,\tilde{I}^2}{\Delta v \,\tau}\, \left\langle \frac{\eta^3}{\sinh^2q}\,(q \,\coth q -1)\right\rangle
\]
where $\quad |x|,|x_0|,\Delta x \ll \tilde \Delta $ and $q=2\eta\Delta x/L$.
So for the one-soliton correlation function we obtain
\begin{equation}
g_1(\Delta x)=4\,\Delta v\,\tau\, \left\langle \frac{\eta^3}{\sinh^2q}\,(q \,\coth q -1)\rangle/\langle \eta \right\rangle^2 , \label{g-1}
\end{equation}
where the averaging in the r.h.s. is performed over the amplitude distribution.
The relative strength of intensity fluctuations for one soliton is given by
\begin{equation}
g_1(0)=\frac{4\,\Delta v \tau}{3}\,\frac{\langle \eta^3 \rangle}{\langle \eta \rangle^2}
\label{fluct-1}
\end{equation}

Next, let us consider a train of $n$ solitons with statistically independent parameters and random, uniform phases. First we consider realizations where exactly $n$-solitons are created.
Then we have
\[
\langle I(x) \rangle = n \langle I_1(x) \rangle = n\,I_1
\]
and
\begin{equation*}
\begin{split}
&\langle I(x) I(x+\Delta x) \rangle  =  n(n-1)\langle I_1(x) \rangle \langle I_1(x+\Delta x) \rangle \\
&+ n \langle I_1(x) I_1(x+\Delta x) + n(n-1) |\langle E_1(x) E^*_1(x+\Delta x) \rangle|^2
\end{split}
\end{equation*}
where $E_1(x)$ is a single-soliton field (\ref{one-sol}) in the real world units. Performing additional averaging over all realizations with different number of solitons as well as over the spatial coordinate $x$ (denoted by an overbar) we arrive at:
\begin{equation}
\begin{split}
g(\Delta x)&=\frac{1}{\langle n \rangle }\,g_1(\Delta x) +\left(\frac{\langle n^2 \rangle}{\langle n \rangle^2} -\frac{1}{\langle n \rangle}\right)
\\
& \times \left(1+\frac{\overline{|\langle E_1(x) E_1^*(x+\Delta x)\rangle|^2}}{I_1^2} \right)
\end{split}
\label{g-n}
\end{equation}
Again if we assume that the typical width of a soliton is much less than its positional support $\tilde \Delta$ one can show that the field correlation term is position independent and is given by
\begin{equation}
\label{field-correl}
\begin{split}
\frac{1}{I_1^2}\,|\langle E_1(x) E_1^*(x+\Delta x)\rangle|^2 & =  \\
\left|\left \langle \frac{\eta}{2\langle \eta \rangle} \, \intop_{-\infty}^\infty \sech[y] \, \sech[y+q]\,e^{-i (q/4\tau \eta^2)\,y} \,dy \right \rangle \right|^2 &
\end{split}
\end{equation}
If additionally $\Delta x \ll \min[4\tau \langle \eta \rangle, L \, \tau^{1/2}]$ (i.e. random phase shift $\Delta \phi =2v \Delta x/L $ can be neglected) the above simplifies to:
\begin{equation}
\frac{1}{I_1^2}\,|\langle E_1(x) E_1^*(x+\Delta x)\rangle|^2 =\left \langle \frac{\eta \, q}{\langle \eta \rangle \, \sinh q} \right \rangle^2
\label{e-e}
\end{equation}

The strength of fluctuations is given by
\begin{equation}
g(0)=\frac{\langle I^2 \rangle}{\langle I \rangle^2}=\frac{4\,\Delta v \tau}{3\langle n \rangle}\,\frac{\langle \eta^3 \rangle}{\langle \eta \rangle^2}+2\,\left(\frac{\langle n^2 \rangle}{\langle n \rangle^2} -\frac{1}{\langle n \rangle}\right)
\label{fluct-n}
\end{equation}

When the argument of the correlation function is large, i.e. the inequality $L/2\Delta x \ll  \sigma_\eta$ holds the function being averaged in Eq.(\ref{g-1}) decays much faster than the PDF $P_\eta$ which can be used to obtain the following asymptote:
\begin{equation}
g_1(\Delta x) \approx \frac{3\zeta(3)}{8} \,\frac{\Delta v \tau P_\eta(0)}{\langle \eta \rangle^2} \,\left( \frac{L}{\Delta x}\right)^4
\label{asymptotes1}
\end{equation}
where $\Delta x \gg L/2 \sigma_\eta$ is assumed. For the full correlation function if one assumes additionally that $\Delta x \gg L \,\max[1/2\sigma_\eta,\sqrt{\tau}]$ the field correlation contribution (\ref{field-correl}) contains the pre-factor $(L/\Delta x)^4$ multiplied by a highly oscillating integral, so its contribution is neglected and one can write down
\begin{equation}
g(\Delta x) \approx g(\infty)+ \frac{g_1(\Delta x)}{\langle n \rangle}
\label{asymptotes2}
\end{equation}
with $g_1(\Delta x)$ given by Eq.(\ref{asymptotes1}) above.

One can see that the correlation function reaches its asymptotic value $g(\infty)=\langle n^2\rangle /\langle n \rangle^2 -1/\langle n \rangle$ following a power law. If the number of emerging solitons follows a Poisson distribution (where the variance is equal to the mean) the limiting value of the correlation function is $g(\infty)=1$ as in the linear case. For the correlation radius $S$ one gets an estimate $S \approx L/2\sigma_\eta$. The latter formula has a transparent physical meaning: the correlation length $S$ is a typical speckle size in any optical system. In the regime considered here bright solitons play the role of ``nonlinear speckles'' so the typical speckle size is the width of a typical soliton, which is given by $L/2\sigma_\eta$.

\end{document}